\documentclass[a4paper,11pt]{article}

\usepackage[utf8]{inputenc}
\usepackage[T1]{fontenc}
\usepackage{graphicx}
\usepackage{hyperref}
\usepackage{xcolor}
\usepackage{authblk}
\usepackage{bm}
\usepackage{amsmath}
\usepackage{amssymb}
\usepackage[english]{babel}

\usepackage{soul} 
\usepackage[normalem]{ulem}

\newcommand{\figref}[2][]{Fig.~\ref{#2}(#1)}
\newcommand{\fullfigref}[2][]{Figure~\ref{#2}#1}
\newcommand{\equref}[2][]{Eq.~\eqref{#2}#1}

\newif\ifshowchanges
\showchangesfalse 

\ifshowchanges
\newcommand{\addrev}[1]{\textcolor{blue}{#1}}
\newcommand{\delrev}[1]{}
\newcommand{\modrev}[2]{\textcolor{blue}{#2}} 

\else
\newcommand{\addrev}[1]{#1}
\newcommand{\delrev}[1]{}
\newcommand{\modrev}[2]{#2}
\fi

\usepackage{flexisym}
\usepackage[font={sf},labelfont=bf]{caption}
\begin{document}

	\title{\modrev{Simulation}{Study} of the \modrev{thermal and acoustic}{acoustic and thermal} response of an elastically anisotropic solid to a \addrev{sub-}nanosecond laser pulse in transient grating spectroscopy}

	\author[1,2]{Jakub Kušnír}
	\author[1]{Tomáš Grabec\thanks{Author to whom any correspondence should be addressed. E-mail: tomas.grabec@it.cas.cz}}
	\author[1]{Petr Sedlák}
	\author[1]{Pavla Stoklasová}
	\author[1]{Hanuš Seiner}

	\affil[1]{Institute of Thermomechanics, Czech Academy of Sciences, Dolejskova 1402/5, 182 00 Prague, Czechia}
	\affil[2]{Faculty of Nuclear Sciences and Physical Engineering, Czech Technical University in Prague, Trojanova 13, 120 00 Prague, Czechia}

	\date{}
	\maketitle

	\begin{abstract}
		Transient grating spectroscopy (TGS) is a material characterization technique based on laser-induced thermoelastic excitation of thermal and acoustic gratings.
        On opaque samples, these gratings are dynamic surface displacements that reflect the sample’s elastic and thermal properties, enabling both types of parameters to be determined from a single experiment.
        Here, we develop a detailed finite element model (FEM) of the TGS experiment that fully captures the coupling between the thermal and mechanical fields, as well as the optical detection of surface displacement using a heterodyning approach.
        Using custom-designed two-dimensional elements, the model is particularly suitable for analyzing TGS measurements on anisotropic media, for which analytical theory is insufficient.
        The simulation captures not only the anisotropic relaxation of the thermoelastic field but also several acoustic features that arise at very short (ultra-transient) timescales and provide additional information about the elastic properties of the examined material.
        The model offers new opportunities for the in silico testing of various modifications of TGS experiments and their applications to a broad class of materials.
	\end{abstract}

	
	\section{Introduction}
	
    Laser-based thermoelastic excitation of acoustic waves was observed only a few years after the laser was developed \cite{White1963}, and the use of spatially periodic excitation followed, where the periodicity was obtained by the interference of two coherent laser beams crossed inside a transparent medium or on the surface of an opaque material \cite{eichler_grating_1977}.
    With further important improvements, such as employing a transmission diffraction mask to create the interfering pump beams from a single pulsed laser beam, which enabled their perfect synchronization, and using a similar optical path for probe beams to create a detection setup with sensitivity enhanced by heterodyning---interferometric amplification of the modulation of the diffracted-laser intensity---this approach gave rise to the method known as transient grating spectroscopy (TGS,\cite{rogers_optical_1997,maznev_optical_1998,stoklasova_laser-ultrasonic_2021}).
	
	In the TGS experiment, the thermoelastic effect leads to the formation of a displacement grating due to the rapid heating of the sample surface. The dynamics of the displacement follows two phenomena: first, a transient thermal grating (TTG) is formed due to thermal expansion. Second, due to the rapidity of the expansion, ultrasonic waves are launched in the sample. 
	Both of these effects depend on the material properties -- thermal and elastic, respectively.
	While the acoustic properties are found from the frequency spectra \cite{maznev_surface_1996,duncan_increase_2016,grabec_guided_2024}, the thermal properties are hidden in the time-domain progress of the signal.
	An analytical theory \cite{kading_transient_1995} provided a fitting formula to extract the thermal diffusivity of the sample from the thermal-grating dynamics, which is commonly used in the literature \cite{johnson_direct_2013,dennett_thermal_2018,choudhry_characterizing_2021}.
	Therefore, joint thermoelastic characterization is feasible by TGS \cite{sermeus_determination_2015,reza_non-contact_2020}.
	
	However, the analytical model involves several simplifications, which is why finite-element models (FEM) have recently been developed to enhance fitting accuracy \cite{simmonds_increased_2024} and to explain the effect of elastic anisotropy on the assessment of thermal diffusivity in single crystals \cite{kusnir_apparent_2023}.
    Molecular-dynamics calculations have also been employed \cite{dennett_bridging_2016} to enable simulation studies of irradiation effects observable by TGS \cite{hofmann_non-contact_2015,reza_thermal_2022}. Nevertheless, all these models target specific limitations of the theoretical fitting approach and do not provide any deeper insight into the TGS experiment itself.
    
    The excitation and detection of the gratings constitute a complex process, combining the fast dynamics of the acoustic response with the much slower yet still dynamic evolution of the thermal and thermoelastic fields, as well as the much faster (immediate) interaction of the surface ripple with the probe beams. The recorded signal is therefore affected by material properties and details of the experimental setup at all these time scales, which calls for a comprehensive numerical model that would allow an in silico replication of the full experiment.

    In this paper, we present such a full finite-element model and show that it is capable of capturing several phenomena observed in the experiment, especially those arising from interactions across the above-mentioned different time scales. Most importantly, the model accurately reproduces the unexpected low-amplitude features in the frequency spectra of TGS signals that arise in measurements on anisotropic materials and often indicate the velocities of bulk longitudinal and transverse waves in the given directions \cite{stoklasova_laser-ultrasonic_2021, zoubkova_ieee_2021}, even though these waves do not contribute to the primary standing-wave patterns generated and detected in the experiment.
    
    As recently discussed in \cite{grabec_utgs_2025}, these so-called ultra-transient features result from the near-field response of the free surface to the pulsed loading and bridge the time scale of the excitation pulse with that of the standing surface acoustic wave (SAW) patterns that dominate the acoustic response. Since the ultra-transient features provide additional information about the material properties, it is essential to understand how their amplitudes depend on various experimental parameters, such as the spatiotemporal characteristics of the excitation pulse, the heterodyne phase of the detection beams, the orientation of the crystallographic cut on which the experiments are performed, and the optical, elastic, thermal expansion, and thermal diffusion coefficients of  the examined material. The developed model enables studying this complex dependence through simulations, serving as a substitute for extensive experimental work.
    
	\section{Methods}
    	\subsection{Transient grating spectroscopy} 
    	\modrev{Transient grating spectroscopy}{TGS} is based on splitting pump and probe beams using a transmission phase grating (phase mask) and employing a 4f imaging system to project and interfere the first-order beams on the sample surface, as illustrated in \figref[a]{fig:TGSSetup}.    	
    	The pump beams create a spatially harmonic interference pattern with a specific period $\lambda$, acting as a spatially narrowband thermoacoustic source.
    	The absorbed laser intensity (and consequently the generated heat) leads to the formation of a thermal grating with the associated surface displacement. The displacement grating connected to the TTG gradually flattens as the temperature homogenizes on the surface.
        The acoustic waves are launched by a sudden expansion, perpendicular to the interference fringes. The counterpropagating waves then interact, creating a standing-wave pattern that manifests as acoustic oscillations of the surface.
        These oscillations are dominated by \modrev{surface acoustic waves (SAW)}{SAW}; however, additional wave modes are also excited by the thermoacoustic source and form complex 'ultra-transient gratings', which can be captured in the experiment \cite{grabec_utgs_2025}.
    	Finally, temperature changes also modulate the surface reflectance according to the material's thermoreflectance coefficient \cite{dennett_thermal_2018}.
    	
    	The time-dependent surface displacement forms a diffraction grating for the probe laser, which is split at the phase mask (similar to the pump laser), and the resulting two probe beams impinge on the sample at symmetric angles. Each incident beam produces reflected and first-order diffracted components (the latter pointing back in the original direction), and vise versa. 
        The signal (modulation of the diffracted component by the surface displacement) is significantly enhanced by heterodyning, which involves combining the diffracted component of one beam with the reflected component of the other beam.
        The phase difference of the two probe beams in the parallel path, called the heterodyne phase $\theta$, determines the character of the signal. 
        At $\theta=0$ or $\pi$, the detection of surface displacement decreases, and the system becomes sensitive only to thermoreflectance.
    	Outside of this phase setting, the signal combines thermoreflectance and surface displacement, with the maximal signal at $\theta=\pm\pi/2$ \cite{johnson_phase-controlled_2012,dennett_thermal_2018}. 
    	
        	\begin{figure}[t]
        		\centering
        		\includegraphics[width=0.75\columnwidth]{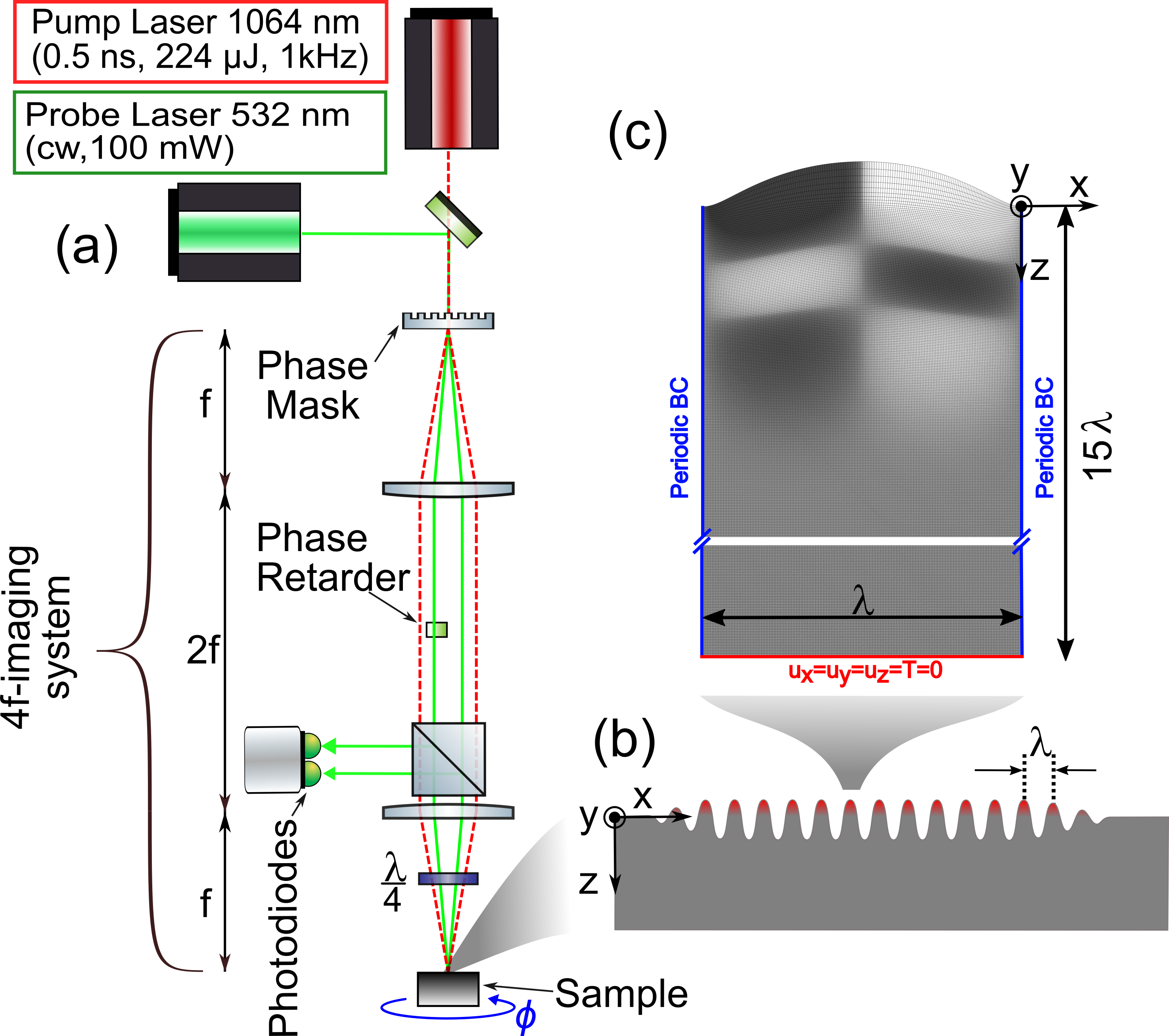}
        		\caption{
        			Simplified schematics of the optical paths in the TGS method (a), illustrating the excited grating at the surface (b), and the correspondence with the computational domain (c).
                    The computational domain is shown with illustrated calculated (largely magnified) displacement, where the shading suggests the $y$-axis displacement magnitude. 
                    }
        		\label{fig:TGSSetup}
        	\end{figure}
	
        	\modrev{The setup used for the measurements shown in this paper was}{Specifics of the TGS setup were}  similar to that presented in previous work by the authors \cite{stoklasova_laser-ultrasonic_2021}, which utilizes the differential heterodyne approach \cite{verstraeten_determination_2015}.
        	The pump beams were provided by an infrared pulsed laser (1064~nm, maximum pulse energy 224~\textmu J, pulse duration (FWHM) 0.53~ns, repetition rate 1~kHz), with the pulse energy lowered in the optical path below 40~\textmu J to remain under the ablation limit (leaving no visible marks on the surface).
        	The probe beams were provided by a green continuous-wave laser (532~nm, nominal power 100~mW, reduced to $\sim 30$~mW on the sample surface). 
        	The TGS setup was configured so that the interference grating wavelength was 10~\textmu m (with the exact wavelength calibrated to $10.064 \pm 0.005$~\textmu m).
        	The signal was then amplified in the bandwidth range of 10~kHz to 1~GHz with a gain of 60~dB and recorded by an oscilloscope card (1.5~GHz bandwidth, vertical resolution of 10~bits, and a sampling rate of 5~GS/s). 
        	To diminish the noise and parasitic signals, the detected signals were averaged 50,000 times in the time domain, and the background signals (acquired with the pump laser blocked) were subtracted.
        	Measurements of angular dispersion were performed by rotating the sample (by the angle denoted in Fig.~\ref{fig:TGSSetup} and hereafter referred to as $\phi$) with respect to the optical axis.

    	\subsection{Simulation of the thermo-mechanical response}
        To simulate the response of the sample to the sub-nanosecond laser pulse, we designed a thermomechanical finite element method (FEM) model as an extension of the quasi-static simulation discussed in previous work by the authors \cite{kusnir_apparent_2023}, where only the surface displacement corresponding to the thermal grating was computed, while the dynamic response (acoustic waves) was omitted.
    	
    	Similarly to the analytical studies of TGS related geometry in Ref.  \cite{kading_transient_1995} and numerical calculations of acoustics using the Ritz-Rayleigh method \cite{stoklasova_forward_2015,grabec_Lamb_2020}, 
    	we do not consider the overall dimensions of the laser beams and the pattern boundaries; instead, we focus on a single interference fringe, as shown in \figref[c]{fig:TGSSetup}.   
    	Moreover, we assume spatial homogeneity along the fringe (along the $y$-axis) -- thus, the computational domain is two-dimensional, with a length of one wavelength ($\lambda$).
    	
    	However, the displacement along the fringes, $u_y$, can be non-zero as a result of elastic anisotropy -- which, in general crystallographic directions, leads to a loss of mirror symmetry with respect to the $xz$. 
        Therefore, the displacement vector must be defined with all three components. 
        To allow this, custom-made elements were designed, and corresponding partial differential equations were formulated for the simulation, which was performed in the COMSOL Multiphysics software:

        The displacement field can be written as
            $u_i(x_j,t)$ and the temperature field as $T(x_j,t)$, where $i=x,y,z$ but $j = x,z$, since the fields are considered homogeneous in the $y$ axis.
                
        The wave equation coupled with thermal expansion can be written as
            \begin{equation}
                \rho \frac{\partial^2 u_i}{\partial t^2}
                - \frac{\partial}{\partial x_j} 
                    \left[ \frac{1}{2} C_{ijkl}  
                        \left(\frac{\partial u_k}{\partial x_l} + \frac{\partial u_l}{\partial x_k}\right)
                    \right]
                + \frac{\partial}{\partial x_j} 
                    \Big[ 
                        C_{ijkl} \beta_{kl} (T - T_0)
                    \Big]
                = 0,
            \label{eq:WaveEq}
            \end{equation}
        with $C_{ijkl}$ being the stiffness tensor, $\rho$ the density, and $\beta_{kl}$ the linear thermal expansion tensor. 

        \addrev{
        The heat diffusion equation in a non-coupled form was considered,
        \begin{equation}
            \frac{\partial T}{\partial t} -
            \frac{\partial}{\partial x_j} \big( \alpha_{ij} \frac{\partial T}{\partial x_i} \big)
            =\frac{P}{\rho c_V},
            \label{eq:HeatEq}
        \end{equation}
        where $\alpha_{ij}$ is the tensor of thermal diffusivity coefficients, $c_p$ is the specific heat capacity, and $P$ is the volumetric heat source representing the absorbed energy of the laser pulse.
        }
        Unlike in the analytical model \cite{kading_transient_1995}, a time-domain Gaussian pulse is considered (instead of a delta-pulse),
        which removes the singularity at the moment of excitation and reduces the temperature peak compared to the instantaneous temporal profile.
        
        \addrev{
            Specifically, the heat source can be prescribed as a thermal pulse along the $x$-direction of one interference fringe with a depth profile in the $z$-direction according to the sample's absorption coefficient $\alpha_z$ in the form
            	\begin{equation}
            		P(x,z,t)\varpropto
            		\left[1-\cos\left(qx\right)\right]
                    \exp\left(-\alpha_{A}z\right)
            		\left[\frac{1}{\sigma\sqrt{2\pi}}
                        \exp\left(-\frac{\left(t-t_{P}\right)^{2}}{2\sigma^{2}}\right)\right]
            		,                
            		\label{eq:thermalPulse}
            	\end{equation}
            where $\sigma$---the standard deviation of the Gaussian function---represents the laser-pulse duration, $t_{P}$ is the time when the pulse reaches maximum intensity (with the initial time shifted to capture the whole Gaussian pulse), and $q=2\pi/\lambda$ is the acoustic wave number. 
        	Taking into account the experimental parameters, the parameters were set as $\sigma=0.25$~ns, $t_{P}=2$~ns, and $\lambda=10$~\textmu m.                
        This formulation of the heat source enables a depth profile given by a finite optical penetration, allowing it to model semi-transparent samples, where the optical penetration depth is comparable to the acoustic wavelength (interference fringe spacing) \cite{zaloznik_analytical_2024}.
        In this work, however, we consider a fully opaque material, where the optical penetration depth is comparable to the mesh size. 
        For such case, to avoid the under-meshing of the source, the heat source can be removed from the right-hand side of Eq. \eqref{eq:HeatEq} and modeled as a Neumann-type surface boundary condition prescribing heat flux (in the same form as in Eq. \eqref{eq:thermalPulse}, except for the dependence on $z$).
        }

        The boundary conditions represent a periodic half-space:
    	Aside from the aforementioned heat flux, the top side was assigned a plane-stress state to mimic the free surface, a periodic boundary was used for the sides of the domain,
    	and the lower boundary was fixed both in displacement and temperature.
        The domain depth of $15\lambda$ was considered for the thermomechanical simulation to minimize the effect of the lower boundary on the surface-guided waves and the reflected waves.
    	Although absorbing boundaries could be used to reduce reflections, they add significant computational cost and are less efficient for general orientations of the wave polarization (displacement) -- which are present when modeling elastic anisotropy and non-principal propagation directions \cite{shen_effective_2015,turkel_local_2023}.

  	    Homogeneous and uniform meshes with three different element sizes were used---25, 50, and 100~nm, corresponding to a total of 9.6, 2.4, and 0.6 million degrees of freedom, respectively (three displacement and temperature fields).
        In all cases, linear elements were used, and the same mesh was used for displacement and thermal fields.
    	A time window of 200~ns was used, with time steps of 0.015, 0.025, and 0.050~ns, respectively, in relation to the element size to fulfill the Courant-Friedrichs-Lewy  conditions (CFL, \cite{courant_partial_1967}).

        \addrev{
            The problem formulation---Eqs. \eqref{eq:WaveEq} and \eqref{eq:HeatEq}---omits the full thermoelastic effect, i.e., the temperature variation owing to the mechanical deformation, i.e., the thermoelastic heat-source term. 
            This simplification is justified by the small strains involved: With the laser-pulse energy being spread over the whole pattern, the expected temperature increase peaks at a few tens of degrees (swiftly dropping down as the heat diffuses), and the expected surface displacement magnitude is in the order of tenths of nanometers \cite{kading_transient_1995,maznev_surface_1999}, leading to strain on the order of $10^{-5}$.
            The formulation dramatically reduces the computational demands, as the heat equation can be solved independently, and the resulting thermal field can then be used as input for the mechanical wave equation.
            }
            For the simulation, the magnitude of the heat source in Eq. \eqref{eq:thermalPulse}---the amount of energy deposited into a material---affects the absolute magnitude of temperature changes and displacement. 
            However, because the thermal expansion coefficient used in our model is linear, the nature of the temperature rise and the displacement characteristics (and resulting frequency spectra) remain arbitrarily scalable.
            Due to the simplified geometry, the simulation aims to reproduce the spatiotemporal character of the displacement dynamics, and the absolute values are not crucial. 
            Thus, the scaling coefficient was chosen so that the maximum temperature peak was on the order of tens of kelvins (expected in the experiment).

       	To compare the simulations with the experimental TGS data, the following single-crystalline nickel material parameters were considered:
        	elastic constants of $c_{11} = 251.9$~GPa, $c_{12} = 157.7$~GPa, and $c_{44} = 119.6$~GPa;
        	density of $\rho = 8.9$~g/cm$^3$;
        	thermal diffusivity coefficient $\alpha = 0.232$~cm$^2$/s;
        	linear thermal expansion $\beta = 13.3\cdot 10^{-6}$~K$^{-1}$;
            \addrev{
                and light absorption coefficient $\alpha_z = 0.139$~nm$^{-1}$
                }
            \cite{stoklasova_laser-ultrasonic_2021,emsley_elements_1998,werner_optical_2009}
            .
        Note that the set of equations allows for more general crystal symmetries to be solved, where the thermal diffusivity and thermal expansion can be anisotropic and more terms of the stiffness tensor independent.
        \addrev{To simulate the angular dispersion, the material tensors were rotated accordingly and the calculation repeated with the transformed tensors (in this case, only the stiffness tensor, as thermal properties are isotropic for cubic materials).}

        	The time evolution of the simulated surface displacement was used to generate the signal in the time-domain, as described in \cite{kusnir_apparent_2023}. 
            The displacement was small with respect to the other dimensions (such as the spacing $\lambda$), as expected in the experiment \cite{maznev_optical_1998}.
        	For the generation of the heterodyne time-domain signal, the variation of the scattered light amplitude depends on the slope of the surface:
            	\begin{equation}
            		\delta A_{\textrm{scatt}}\left(x,t\right)\varpropto\frac{\partial u_{z}(x,z=0,t)}{\partial x},
                    \label{eq:ScattLightAmplitude}
            	\end{equation}
        	\delrev{where $\phi$ denotes the measured angle of rotation of the sample. } 
        	The amplitude of the diffracted beam in the direction of incidence over the domain is given as
        	   \begin{equation}
            	A_{\textrm{diff}}(t)=\intop_{0}^{\lambda}\delta A_{\textrm{scatt}}(x,t)\exp(iqx)\textrm{d}x.
        	   \end{equation}
            The reference signal amplitude for the simulation is chosen as a hundredfold amplified diffraction amplitude at the time when the maximum temperature of the sample 
                \delrev{in the first angle $\phi_{0}$ considered }
                is reached:
            	\begin{equation}
            		A_{\textrm{ref}}=100 \, A_{\textrm{diff}}(t_{\textrm{Tmax}}).
            	\end{equation}
        	
            Similar to the calculated diffracted signal, the reference signal is also a complex number -- and the relative difference of their angle sets the heterodyne phase $\theta$. 
            To maximize the influence of surface displacement on the resulting signal,
            $A_{\textrm{diff}}$ and $A_{\textrm{ref}}$ are set to be in phase, mimicking the experimental conditions of $\theta=\pi/2$.
        	Then, the generated heterodyne time-domain signal intensity is given as a perturbation of the reference signal by the diffraction signal,
        	\begin{equation}
        		I_{\textrm{gen}}(t)=
        		\left|A_{\textrm{ref}}+A_{\textrm{diff}}(t)\right|^{2}-\left|A_{\textrm{ref}}\right|^{2}.
        	\end{equation}

	\section{Results and Discussion}
        \subsection{Example of simulated displacement and temperature fields}    
\begin{figure}
        		\centering
        		\includegraphics[width=1\linewidth]{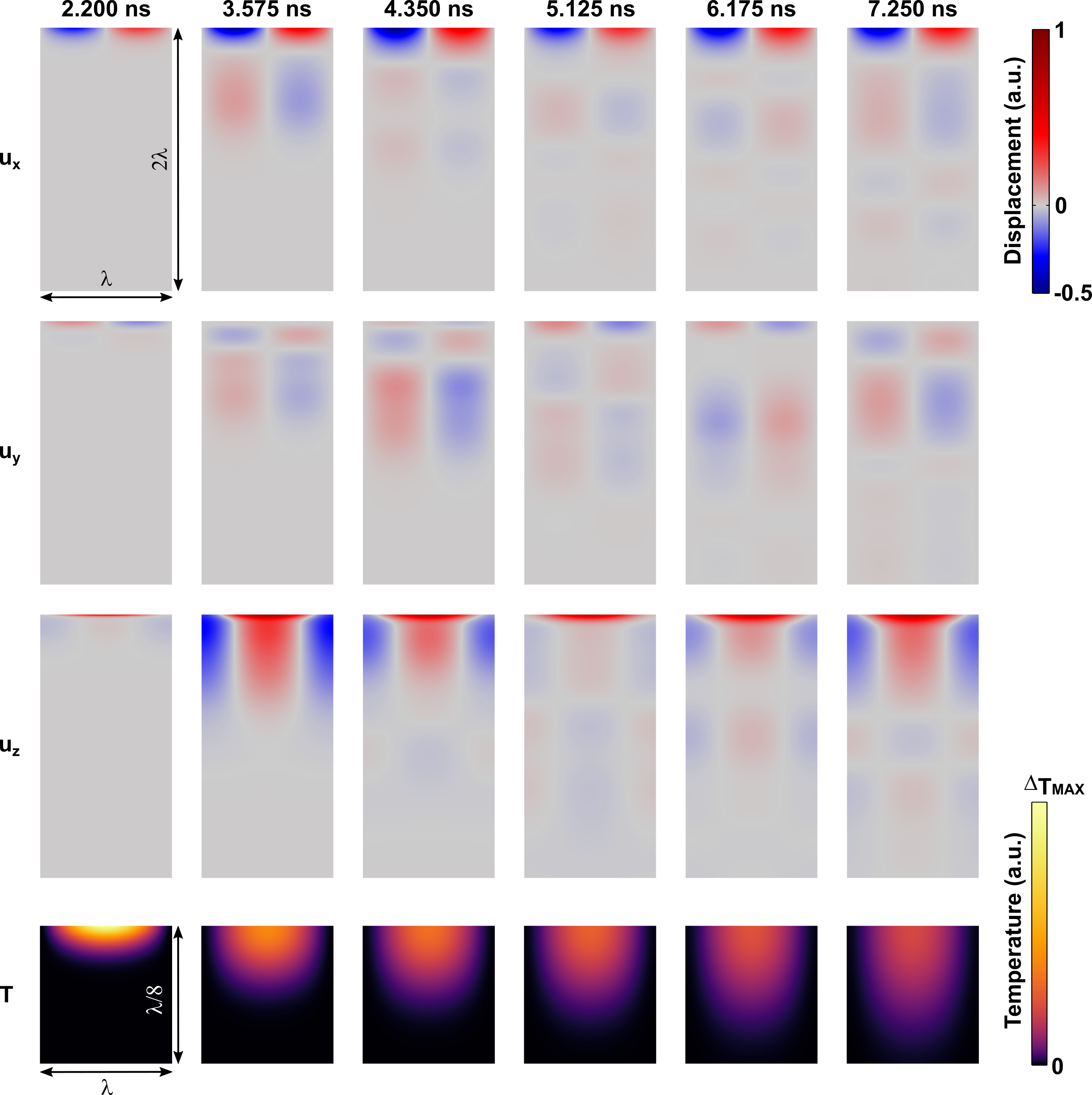}
        		\caption{
                    Simulated displacement fields ($u_x$,$u_y$, and $u_z$) for the direction 30$^\circ$ off the [001] in Ni(110) at various time steps: reaching the maximum surface temperature (2.2~ns, note that the thermal pulse, \equref[]{eq:thermalPulse}, is centered around $t_P = 2$~ns), followed by 
                    specific times of the SAW oscillation.
                    Below are the corresponding temperature fields showing the gradual homogenization at the surface and its spread towards the depth.
                    Note the different scales in $x$ and $z$ direction, and also between the displacement and temperature fields.
        		}
        		\label{fig:Fields}
        	   \end{figure}

        The time evolution of the simulated displacement and temperature fields for the direction 30$^\circ$ off the [001] in Ni(110) is shown in Fig.~\ref{fig:Fields}. 
        
        Although the simulated thermal flux is concentrated at the surface, the heat spreads to a depth of hundreds of nanometers even during the duration of the thermal pulse itself.
        Therefore, the initial spike in surface temperature is quickly diminished as the heat spreads to depth and homogenizes over the surface.  

        The displacement fields document the necessity of calculating all three displacement components when a general direction is considered. The displacement fields in the propagation direction ($u_x$) and the in-plane perpendicular direction ($u_y$) show a harmonic character along the surface and document the propagation of acoustic waves to the depth of the domain.
        The out-of-plane displacement ($u_z$) is strongly coupled to the temperature field, as the free surface is the dominant direction for thermal expansion. Therefore, a "bulging" of the surface---the thermal grating---is formed, on which the acoustic oscillation is superposed (as shown in \figref[c]{fig:TDSignal}).
        
    	\subsection{Time-domain signal and thermal diffusivity}
    	To validate the simulation, in \fullfigref[]{fig:TDSignal}, we compare the simulation-generated signal $I_{\textrm{gen}}$ (calculated with the element size of 25~nm) with the experimentally measured $I_{\textrm{exp}}$ on a single-crystalline sample of nickel in the orientation (110)[001].
    
        	\begin{figure}[t]
        		\centering
        		\includegraphics[width=0.85\columnwidth]{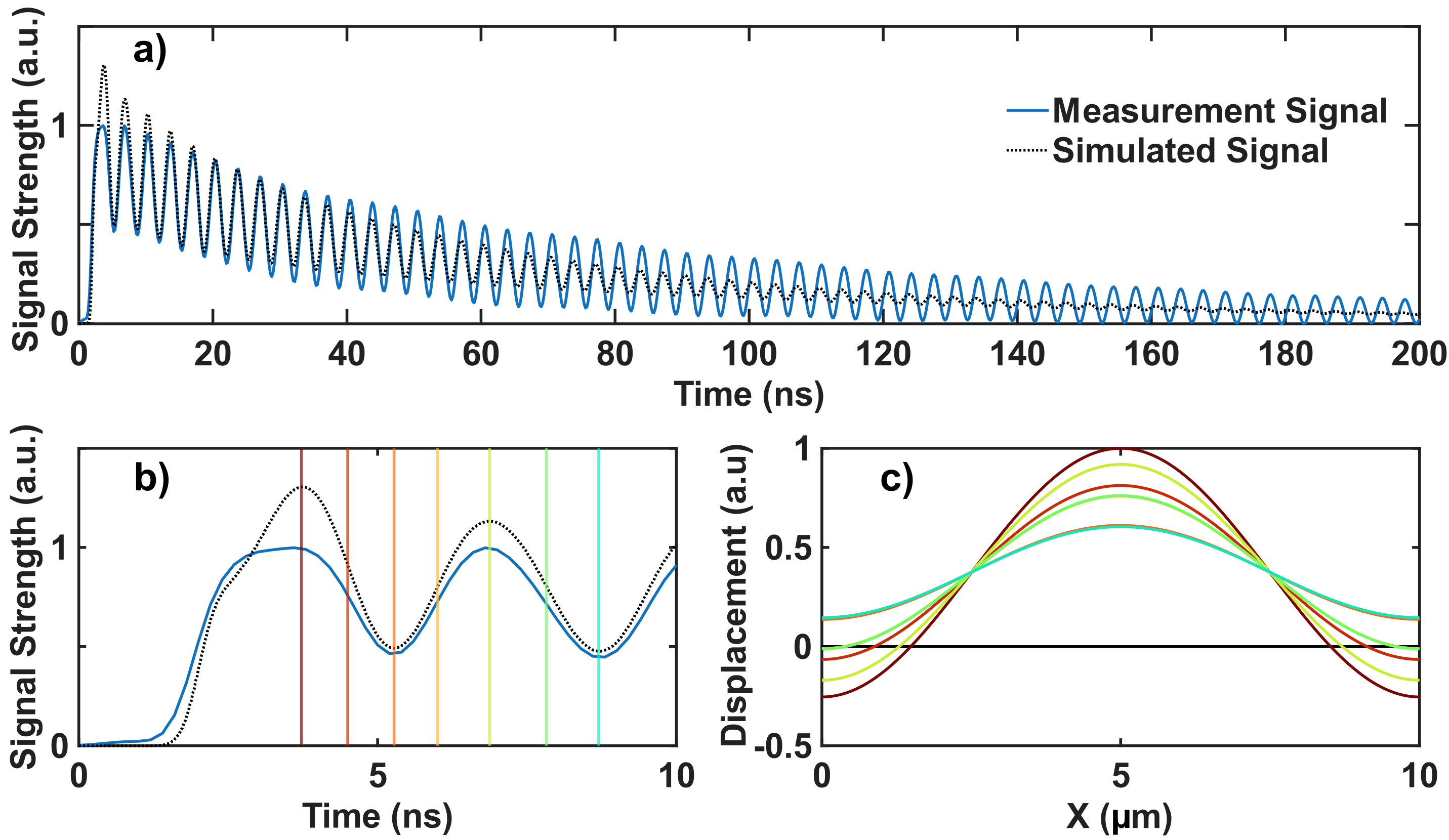}
        		\caption{
        			(a) Comparison of experimental (blue) and simulated (black) time-domain signals obtained from measurement on Ni(110)[001] at the acoustic wavelength $\lambda = 10$~\textmu m calculated for the element size of 25~nm
        			(b) Close-up on the first 10 nanoseconds after the pump pulse.
        			(c) Simulated surface displacement profiles at given time steps marked in (b).
        		}
        		\label{fig:TDSignal}
        	\end{figure}
    	
    	Both signals exhibit similar temporal evolutions.
    	Their basic shape---slow decay superposed with oscillations---is also documented by the calculated surface-displacement profiles (see \figref[c]{fig:TDSignal}): Thermal expansion forms high fringes due to the thermal grating, which monotonously decrease as the temperature homogenizes along the $x$ direction (within the interference pattern). 
    	The acoustic oscillations are superimposed over the thermal grating, their amplitude being below that of the original thermal peak.
    	Note that the relative amplitudes and durations of acoustic oscillations and thermal peaks are material-specific.
    	
       	The agreement between the simulated and measured signals is promising, considering the significantly simplified geometry of the simulation. 
    	There are two main differences -- the overall profile near the beginning of the signal and the attenuation of acoustic oscillation.
    	
    	The first discrepancy is documented in \figref[b]{fig:TDSignal} showing the beginning of the signal.
    	The most probable reason for this discrepancy is the omission of the thermoreflectance effect in the simulation: As the sample surface is heated rapidly in the initial nanoseconds of the measurement, its surface reflectance changes with temperature, 
        modulating the strength of the diffracted beam.
        This modulation decreases exponentially with time -- and typically faster than the surface displacement, since the temperature spike at the surface drops faster than the displacement, as discussed in literature \cite{kading_transient_1995, johnson_phase-controlled_2012} and illustrated in Fig.~\ref{fig:Fields}.
        That agrees with the time-profile of this discrepancy, as it is present in the first few nanoseconds after the excitation, with the signals coming significantly closer afterward.
        \addrev{
            Note that including the thermal variation of reflectivity would be technically possible. 
            It would require adding a necessary material parameter and evaluating the reflectivity at each time step, to evaluate its influence on the scattered-light amplitude (Eq. \eqref{eq:ScattLightAmplitude}).
            However, it would necessitate a precise (sample-dependent) scaling of the heat source to evaluate the exact change in temperature.
            Due to the focus on the dynamics of the response and the resulting signal, the simulation is kept simple and arbitrarily scalable with respect to the exact amount of energy (heat) put into the sample. 
        }
    	
\begin{figure}
        		\centering
        		\includegraphics[width=0.85\linewidth]{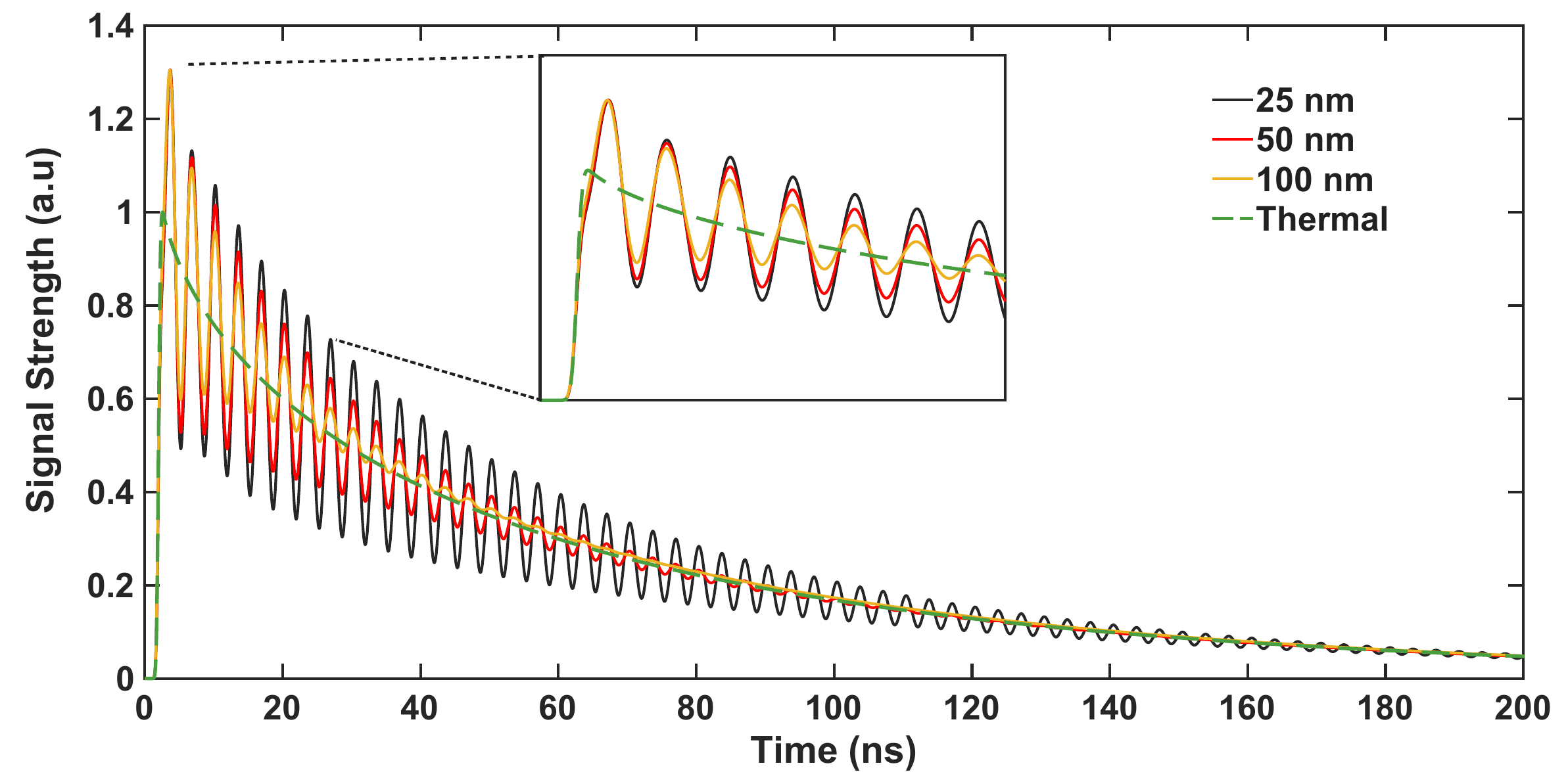}
        		\caption{
        			Simulated time-domain signals comparing attenuation (lifetime) of the acoustic oscillations with mesh sizes of 25, 50, and 100~nm shown in black, red, and yellow, respectively.
        			While the acoustic lifetime differs significantly, thermal profile agrees with the quasi-static simulation omitting the acoustic dynamics (denoted in green).   
        		}
        		\label{fig:simulatedTDSignals_meshSize}
        	\end{figure}

    	The second difference is the duration of the acoustic oscillations:
    	In the experiment, the lifetime of the surface acoustic waves is determined by the finite dimensions of the excitation pattern, within which the amplitude gradually decreases as the laser-beam intensity diminishes towards the edges of the pattern (assuming a Gaussian profile of energy),
        material damping and scattering \cite{grabec_surface_2022},
        and scattering on the surface roughness \cite{sarris_attenuation_2021}.
    	However, none of these effects are present in the simulation, as the periodic boundaries in the $x$-direction and homogeneity in the $y$-direction lead to an infinite pattern.
    	Thus, we attribute the attenuation to numerical mesh scattering.
    	This is confirmed in Fig.~\ref{fig:simulatedTDSignals_meshSize} by a comparison of the simulated signals with different element sizes -- where the coarser the mesh, the larger the numerical attenuation \cite{marfurt_accuracy_1984}.
    	Thermal decay is well captured in all of these mesh sizes, as confirmed by the comparison with the quasi-static simulation \cite{kusnir_apparent_2023}. 
        The mesh-scattering influence is also confirmed by the agreement of the signals at the very beginning of the acoustic oscillations (shown in the inset of  Fig.~\ref{fig:simulatedTDSignals_meshSize}).

    	\subsection{Frequency-domain signal comparison}
\begin{figure}[!t]
        		\centering
        		\includegraphics[width=1\columnwidth]{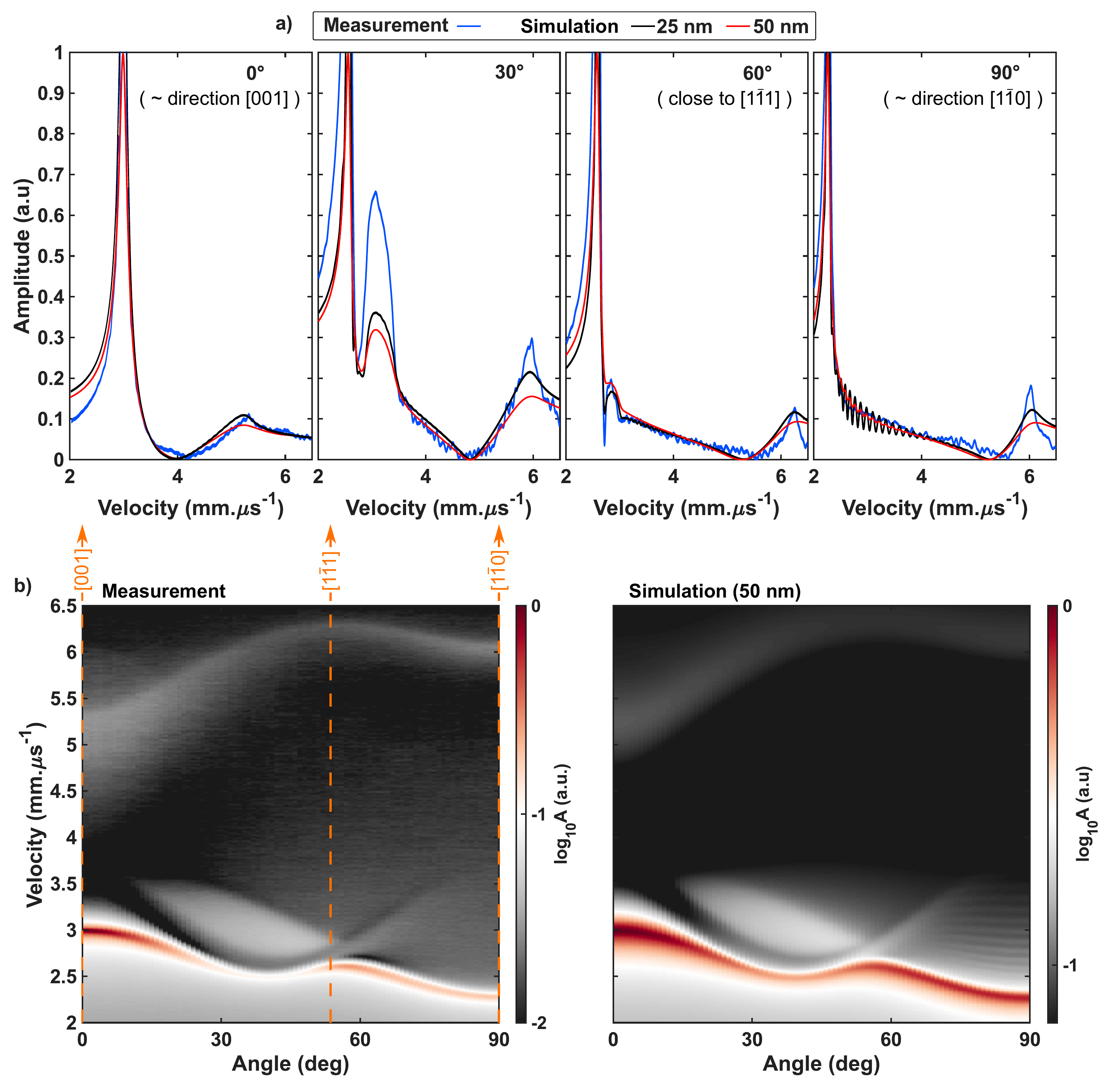}
        		\caption{
        			(a) \addrev{Frequency-domain (or rather, velocity-domain) signals } of signals measured by TGS (blue) and obtained by FEM with the mesh size of 25~nm (black) and 50~nm (red), indirections of 0$^\circ$, 30$^\circ$, 60$^\circ$, and 90$^\circ$ off the direction [001].
                    The amplitudes were adjusted for the best fit of the low-amplitude features.
                    (b) \modrev{Frequency-angular}{Velocity-domain angular-resolved} maps measured by TGS and obtained by FEM on the left and right, respectively. 
                    Note that the simulated map was obtained with the 50nm mesh for the sake of computational demands.
                    }
        		\label{fig:Maps}
        	\end{figure}

    	Although the time-domain signal is useful for determining the thermal diffusivity coefficient, the elastic properties are characterized by the frequency spectra \addrev{(or velocity spectra, with $v=\lambda f$)} of the acoustic oscillations.
    	As shown in \figref[a]{fig:Maps}, the simulated and measured frequency spectra agree very well \addrev{in a variety of angles (propagation directions)} -- to the extent of capturing not only the main peak of the surface acoustic wave but also other acoustic features originating in the ultra-transient grating phenomena \cite{grabec_utgs_2025}. 
    	These features present a minute modulation of the surface displacement during the first several nanoseconds of the acoustic signal -- that is, with significantly lower prominence and orders-of-magnitude shorter timescale than the SAW. 
        Yet, they are captured very well in the simulation
           \delrev{ , as shown in Fig.~\ref{fig:Maps} }
            -- the agreement with the measured signals documents that the surface displacement and its time profile is realistic from the beginning of the simulation, where the finite duration of the pulse plays the largest role.
    	Capturing the ultra-transient effects in the simulation with excellent agreement to the experiment nominates the FEM simulation for further studies of these effects.
        
        \modrev{
            For a further comparison with the experimental results, 
        	we performed a calculation of the frequency-angular dispersion on the considered crystallographic cut (110) with a fine step of 1$^\circ$.
           	These frequency-angular maps are shown in \figref[b]{fig:Maps}.
            }
            {
            For further comparison of the experimental and computational results, 
            we measured and calculated angular dispersion of the considered crystallographic cut (110) with a fine step of 1$^\circ$,
            with the resulting frequency-domain angular-resolved maps shown in \figref[b]{fig:Maps}
            -- again, with frequency recalculated to velocity to avoid the dependence on chosen wavelength $\lambda$.
            }
    	Note that, for the sake of computational demands, we performed the calculation with the 50nm mesh -- while the 25nm mesh provides near-perfect results, it is also nearly seven times more computationally demanding (due to the increase in the number of elements and lower time step).
    	The agreement of these maps demonstrates the ability of the designed simulation to capture the surface dynamics including ultra-transient phenomena in various studied crystallographic directions, where the wave polarizations are coupled
            \addrev{(as the rotated stiffness tensor follows less symmetries in general directions).}

    	Note that the finite depth of the computational domain can cause regular stripe artifacts at frequencies above the main SAW peak, which can be observed simulated map in \figref[b]{fig:Maps} between angles 60$^\circ$ and 90$^\circ$, where the acoustic properties allow whole-domain thickness resonances.
        As documented by the comparison of 25nm and 50nm mesh in \figref[a]{fig:Maps}, the lower mesh scattering allows more pronounced resonances of this kind. 
        
    \section{Conclusion}
	We developed and validated a thermomechanical finite element model of transient grating spectroscopy that captures the coupled thermal and acoustic response to a nanosecond laser pulse in an elastically anisotropic solid. 
	The model employs a periodic two-dimensional domain with three components of displacement. 
    \modrev{The excitation is modeled as surface-concentrated heat flux following a Gaussian pulse with a harmonic spatial distribution.}{The excitation is modeled as a volumetric heat source following a Gaussian pulse with a harmonic spatial and depth-dependent profile, reducible to a surface heat flux for opaque materials.}
    The resulting simulated out-of-plane surface displacement is then converted to a heterodyned signal, equivalent to that measured by TGS.
	
	Quantitative comparison with measurements on Ni(110) shows good agreement in the time domain, including the slowly decaying transient thermal grating envelope modulated by surface acoustic oscillations.
    Frequency spectra of the simulated signals agree very closely with the experimental results, including not only the main peak of the surface acoustic wave but also features corresponding to surface-skimming waves and other acoustic phenomena at the surface. 
    This agreement is confirmed by a \modrev{frequency-angular dispersion}{frequency-domain angular-resolved} map, which shows that the simulation captures even the minute and ultra-transient acoustic responses of the surface, which are present only shortly after the excitation.
	
	These results indicate that the presented FEM framework provides a reliable and computationally efficient tool for simulating TGS signals in anisotropic materials, enabling the joint interpretation of thermal diffusivity and acoustic dispersion from a single experiment.
	The approach offers a basis for systematic studies of crystallographic anisotropy and the dispersion of guided modes, as well as the effects of experimental-setup modifications on the resulting signals and the strength of various acoustic modes in the resulting signal.

	\section*{Acknowledgements}
        This work was financially supported by the Czech Science Foundation [Project No. 22-13462S], Operational Program Johannes Amos Comenius of the Ministry of Education, Youth and Sport of the Czech Republic, within the frame of project Ferroic Multifunctionalities (FerrMion) [Project No. CZ.02.01.01/00/22\_008/0004591], co-funded by the European Union, and by the Grant Agency of the Czech Technical University [No. SGS25/168/OHK4/3T/14].
	
	
	\section*{Author contributions}
		Conceptualization: JK, TG, PSt, PSe, HS.
        Data Curation: JK.
        Formal analysis: JK, TG.
        Investigation: JK, TG.
    	Methodology: JK, TG, PSt, PSe.
		Software: JK, TG.
        Supervision: PSt, PSe, HS.
		Writing—Original Draft: JK, TG.
		Writing—Review and Editing: PSe, HS.
		Visualization: JK.
		Funding acquisition: PSe, HS.
	
	
	\section*{Data availability}
		The data that support the findings of this study are openly available at the following URL/DOI: \\ https://doi.org/10.5281/zenodo.17724594.

	\bibliographystyle{unsrt}
	\bibliography{Bib_FEMofTGS_Kusnir2025}

\end{document}